\documentclass[aps,prb]{revtex4} 

\usepackage{graphicx}
\usepackage{dcolumn}
\usepackage{bm}

\newcommand{\comment}[1]{}

\begin{document}


\title{Ten Equations that Shook the Quantum World:
Bose-Einstein Condensation, Superfluidity,
and the Quantum-Classical Transition}


\author{Phil Attard}
\affiliation{
{\tt phil.attard1@gmail.com}}


\begin{abstract}
The transition from the quantum to the classical world
and its relation to Bose-Einstein condensation
and superfluidity is explained in ten equations.
\end{abstract}


\maketitle


%
\section{Quantum-Classical Transition}
\setcounter{equation}{0} \setcounter{subsubsection}{0}
%

\subsection{Particle Position and Momentum}

A particle has position and momentum,
and the evolution of these in time defines its trajectory.
To many this is self-evident,
a simple extrapolation of what they see every day with their own eyes.
But to others,
namely those with some familiarity with quantum mechanics,
it is just plain wrong,
a direct contradiction to the conventional understanding of the subject.
To most quantum mechanicians,
the wave aspects of a particle mean, amongst other things,
that it is spread out in space,
that it interferes with other particles,
and that it does not posses at any instant
a definite position and momentum.

There is compelling evidence to support
the wavy nature of particles,
including the original observations of the interference patterns
in electrons scattered from crystals
and in the double slit experiment.
Going beyond wave attributes,
the wide-spread belief that particles do not have both position and momentum
is an extrapolation of Heisenberg's uncertainty principle:
for any wave function
the variance in the momentum operator
(ie.\ the expectation value of the square 
less the square of expectation value)
times the variance of the position operator
cannot be less than $\hbar^2/4$,
where $\hbar$ is Planck's constant divided by $2\pi$.
This is an exact mathematical result about which there is no dispute.
However, any physical interpretation beyond this
requires a dependent chain of assertions,
of which the three most important are:
(1) that an expectation value is a physical measurement,
(2) that the product of the variances
is the same as the product of the measurement uncertainties,
and (3) that the inability to measure these with infinite precision
implies that the particle cannot possess them simultaneously.
Each one of these assertions may be disputed.
I invite the reader to reflect upon
the distance between the mathematical result
that is Heisenberg's uncertainty principle
and the physical interpretation that particles cannot have
position and momentum simultaneously.

A concrete counter-example suffices.
The de Boglie-Bohm pilot wave theory
(Bohm 1952, de Broglie 1928)
ascribes to particles position and momentum simultaneously,
and yet it reproduces all of the known mathematical results
of quantum mechanics,
including the Heisenberg uncertainty relation
(Goldstein 2024).
Therefore it is a logical contradiction to assert
that the equations of quantum mechanics imply that
a particle cannot possess simultaneously position and momentum.

Whether or not particles have position and momentum
is an important conceptual and practical issue.
In the classical world the positions and momenta of all the particles
in the system, a point in phase space,
specifies the state of the system,
and Newton's or Hamilton's equations of motion
create a trajectory in phase space
that gives the system's evolution over time.
In contrast, in the quantum world
the wave function specifies the state of the system,
and the time evolution of the wave function
is given by Schr\"odinger's equation.
Because almost all quantum mechanicians
deny that particles have both position and momentum,
little attention has been paid to the possibility
of formulating the evolution of a quantum system
in terms other than Schr\"odinger's equation.

And yet there are compelling arguments
for exploring alternative possibilities.
On the one hand, classical mechanics must be valid in the classical regime,
which, roughly speaking, is what we observe with our own eyes.
After all, Newton's and Hamilton's equations of motion
are empirical equations based on measured data.
On the other hand, quantum mechanics is the fundamental theory,
and the equations of quantum mechanics
must therefore determine those of classical mechanics.
The empirical evidence
(eg.\ the $\lambda$-transition in liquid helium)
is that the quantum-classical transition is continuous.
This means that
there must be remnants of classical behavior in the quantum system.
These considerations,
together with the above arguments for particles possessing
position and momentum,
suggest that, at least in the transition region,
there must be an alternative to Schr\"odinger's equation
that lies closer to Newton's and Hamilton's equations of motion
for the time evolution of the system.
It is not inconceivable that
the validity of such an alternative formulation
might extend beyond the transition region
deep into the quantum regime itself.

In this essay we introduce particle position and momentum
not by pilot wave theory,
but rather via the momentum eigenfunction,
\begin{equation}
\zeta_{\bf p}({\bf q})
=
\prod_{j=1}^N V^{-1/2} e^{-{\bf p}_j\cdot{\bf q}_j/\mathrm{i}\hbar}
\equiv
\prod_{j=1}^N \zeta_{{\bf p}_j}({\bf q}_j).
\end{equation}
Here the cubic subsystem has volume $V=L^3$,
and the spacing of the momentum states is $\Delta_p = 2\pi\hbar/L$.
The periodicity of the eigenfunction
ensures that the momentum operator is Hermitian.
The orthonormal eigenfunctions,
$\langle \zeta_{{\bf p}'} | \zeta_{{\bf p}''} \rangle
= \delta_{{\bf p}',{\bf p}''}$,
which are not yet symmetrized,
consist of Fourier factors
that  reflect the wave attributes of the particles.
When we speak of particle $j$ having position ${\bf q}_j$
and momentum ${\bf p}_j$,
what we really mean is that the particle is at that position
in that momentum eigenstate,
and that associated with the particle is the complex number
$\zeta_{{\bf p}_j}({\bf q}_j)$.

\subsection{Superposition of States}

Before discussing an alternative
for the time evolution of the quantum system based
on the particles' positions and momenta,
some other differences between quantum
and classical systems are raised
with a view to understanding the transition from one to the other.

First is the superposition of quantum states,
which arises because the sum of allowed wave functions
is also an allowed wave function.
Alternatively,
Schr\"odinger's equation is linear,
and so the time evolution of a sum of wave functions
is the sum of the time evolution of the individual wave functions.

In the classical world,
we cannot add particle configurations
(ie.\ points in phase space)
since this would change particle number,
or cause physically impossible particle overlap,
or multiple positions for each particle.
In the classical world a particle has one, and only one,
momentum at a time.

In the popular imagination, Schr\"odinger's cat
exemplifies quantum superposition.
It is surprising how many otherwise intelligent scientists
take Schr\"odinger's cat literally rather than metaphorically.
It is usually forgotten that Schr\"odinger invented his cat
to argue against the completeness of quantum mechanics;
he believed that the proposition that an unobserved cat
was both dead and alive
so nonsensical that no one would mistake it for physical reality.

Whether or not Schr\"odinger and Einstein were correct
in arguing that quantum mechanics is incomplete,
they were certainly correct in pointing out that the macroscopic objects
of the classical world do not exist in superposition.
But since, as we have said above,
quantum mechanics underlies classical mechanics,
the question is:
how precisely does the superposition of states
disappear in the quantum to classical transition?

To make the discussion more concrete,
in the present context that focusses on momentum eigenfunctions,
the superposition of two of these would read
\begin{equation}
\psi({\bf q})
=
\frac{1}{\surd 2}
\left[
\zeta_{{\bf p}'}({\bf q})+ \zeta_{{\bf p}''}({\bf q})
\right].
\end{equation}
Here ${\bf p}' \ne {\bf p}''$ are two different momentum configurations
of the system, and the factor of $1/\surd 2$ ensures
that the wave function is normalized.
This is a valid wave function.
However classically it is not permitted for boson $j$ at ${\bf q}_j$
to have different momenta ${\bf p}'_j$ and ${\bf p}_j''$ simultaneously.

In order to understand the suppression of superposition
in classical systems
we have to introduce the concept of a subsystem.
This is the specific part of the universe
in which we are really interested
and that is our primary focus.
But we cannot entirely neglect the rest of the total system
with which it can exchange energy.
Here that remainder is called the reservoir,
although it is just as well called the environment or the heat bath.
The subsystem is sometimes called an open quantum system.

\begin{figure}[t]
\centerline{ \resizebox{8cm}{!}{ \includegraphics*{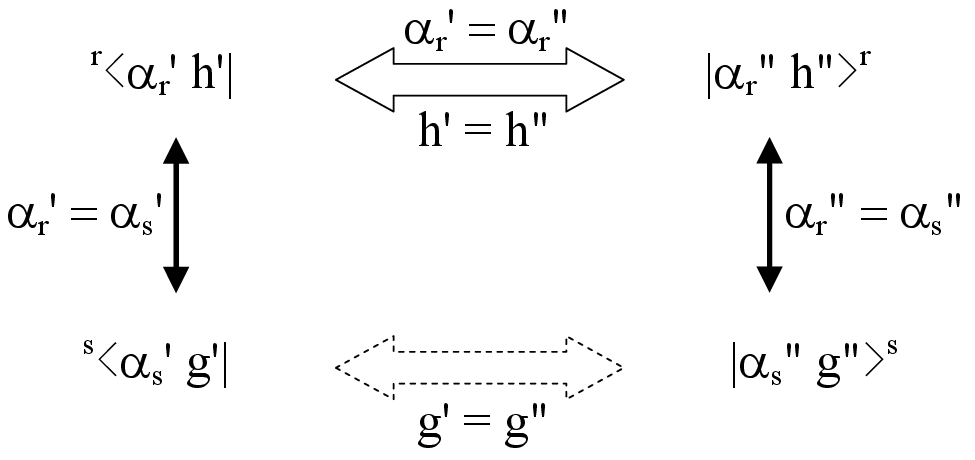} } }
\caption{\label{Fig:supp}
Collapse into decoherence due to entanglement.
The vertical arrows represent energy conservation
causing the one-to-one relationship between the principle energy states
of the subsystem and reservoir.
The solid horizontal arrow represents the orthogonality
of the reservoir energy eigenfunctions.
The dashed horizontal arrow represents
the cancelation due to the random phases
of the degenerate subsystem energy states.
}
\end{figure}

Because energy is conserved in the total system,
the total wave function is the  superposition of entangled products
of the subsystem and the reservoir energy eigenfunctions,
$\psi({\bf q}^\mathrm{s},{\bf q}^\mathrm{r})
=\sum_{\alpha;g,h} C_{\alpha;g,h}
\phi^{E,\mathrm{s}}_{\alpha g}({\bf q}^\mathrm{s})
\phi^{E,\mathrm{r}}_{\alpha h}({\bf q}^\mathrm{r})$,
with $E^\mathrm{s}_\alpha + E^\mathrm{r}_\alpha = E^\mathrm{total}$.
The expectation value of an operator on the subsystem
consists of a sum over eight indeces:
two principle and two degenerate energy states
on each side of the inner product of the wave functions of the subsystem
and of the reservoir (Fig.~\ref{Fig:supp}).
Because of the orthogonality of the reservoir wave functions,
the double sum over principle energy states
of the inner product of reservoir wave functions collapses
into a single sum, which means that
the reservoir can exist in only one principle energy state at a time.
But by energy conservation a unique reservoir energy implies
a corresponding unique subsystem energy,
which means that the subsystem must also have collapsed
into a decoherent mixture of pure principle energy states.
The degenerate energy states of the subsystem likewise collapse
into a mixture of pure states
because they have random phases
and any superposition of them cancels upon summation.
The degenerate energy states of the reservoir
are decoherent by orthogonality,
and the sum over them yields the exponential of the reservoir entropy,
which is just the Maxwell-Boltzmann factor
that weights the principle energy states of the subsystem.

In summary, a subsystem entangled with its environment
collapses into a decoherent mixture of pure states
with quantum superposition suppressed.
The states are weighted by the Maxwell-Boltzmann factor,
which represents the entropy of the reservoir.
The relevance of this to the transition to the classical world
is that macroscopic objects may be subdivided into microscopic subsystems,
each entangled with its surroundings
and therefore decoherent (ie.\ no superposition allowed).

It can be mentioned that entanglement with the environment
as the source of decoherence was originally raised
in the context of quantum measurement theory
(Joos and Zeh 1985, Schlosshauer 2005, Zurek 1991).
The mathematical details of the present
statistical thermodynamic treatment
may be found in Attard (2018, 2021).

The suppression of superposed wave states in an open quantum system
leads to quantum statistical mechanics,
which gives the statistical average of a quantum operator
as the sum over a mixture of pure energy states.
This has the generic form
of the trace of the product of the operator
and the Maxwell-Boltzmann probability operator
(also known as the density operator).
A trace is a scalar, which holds in any basis,
including that of the momentum eigenfunctions, Eq.~(1).
These represent the trace
as the sum over momentum states and the integral over positions
of the subsystem,
which is akin to the integral over phase space
that defines classical statistical mechanics.
The total entropy,
which contains all of the thermodynamic properties of the system,
is the logarithm of the weight of configurations.
The latter is the grand partition function,
and for bosons it can be written
\begin{equation}
\Xi^+ =
\sum_{N=0}^\infty
\frac{z^N}{V^N N!}
\sum_{\bf p} \int \mathrm{d}{\bf q}\;
e^{-\beta {\cal H}({\bf \Gamma})} \omega({\bf \Gamma})
\eta^+({\bf \Gamma}) .
\end{equation}
Here a point in quantized phase space is a configuration of the subsystem,
${\bf \Gamma} = \{ {\bf q},{\bf p}\}$,
the fugacity is $z=e^{\beta \mu}$,
and the inverse temperature of the thermal reservoir or environment
is $\beta = 1/k_\mathrm{B} T$.
The commutation function,
$ \omega({\bf \Gamma})
\equiv
e^{\beta {\cal H}({\bf \Gamma})}
e^{{\bf p}\cdot{\bf q}/\mathrm{i}\hbar}
e^{-\beta \hat{\cal H}({\bf q})}
e^{-{\bf p}\cdot{\bf q}/\mathrm{i}\hbar}$,
is a short-ranged function that accounts for the non-commutativity
of the position and momentum operators.
The symmetrization function,
$\eta^+({\bf \Gamma})
\equiv
\sum_{\hat{\mathrm P}}
e^{-[{\bf p}-\hat{\mathrm P}{\bf p}]\cdot{\bf q}/\mathrm{i}\hbar}$,
which is a sum over all permutations of the  ratio
of the original to the permuted eigenfunction,
arises from the symmetrization of the wave function,
which is now discussed.

\subsection{Symmetrization and Condensation}

It is a fundamental requirement of quantum mechanics
that the wave function remain unchanged
upon the interchange of two identical bosons,
and that it changes sign upon the interchange of two identical fermions.
More generally the wave function is said to have even or odd
permutation symmetry, respectively.
There is no analogue of this in the classical regime,
but nevertheless it has ramifications in the region of the quantum-classical
transition.

A well-known consequence of wave function symmetrization
is that no two fermions may occupy the same one-particle state,
which is why no more than two electrons
can be in any hydrogen-like electronic orbital,
and if there are two then they must have opposite spin.
Conversely, an unlimited number of bosons may occupy
the same one-particle quantum state
such as a momentum state.

This property of bosons gives rise to
the phenomenon of Bose-Einstein condensation,
which is driven by occupation entropy, as is now explained.
We focus on bosons and on momentum eigenfunctions,
which we symmetrize by summing equally over all permutations,
\begin{equation}
\zeta^+_{\bf p}({\bf q})
=
\frac {1}{\sqrt{N! \chi^+_{\bf p}}}
\sum_{\hat{\mathrm P}}
\zeta_{\hat{\mathrm P}{\bf p}}({\bf q}) ,
\;\;
\chi^+_{\bf p}
\equiv \sum_{\hat{\mathrm P}}
\langle \zeta_{{\bf p}}| \zeta_{\hat{\mathrm P}{\bf p}} \rangle
= \prod_{\bf a} N_{\bf a}! .
\end{equation}
Here $\hat{\mathrm P}$ is the permutation operator,
and the occupation number,
$N_{\bf a} = \sum_{j=1}^N \delta_{{\bf p}_j,{\bf a}}$,
is the number of bosons in the one-particle momentum state ${\bf a}$.
The symmetrization factor, $\chi^+_{\bf p}$,
is required to normalize the symmetrized wave function;
as the number of non-zero permutations (explained next),
which depends upon the occupancies,
its logarithm may be called the occupation entropy.
As can be seen in Eq.~(3), the partition function,
whose logarithm gives the total entropy,
includes a sum over permutations.
This entropy increases with increasing occupation of the momentum states,
which explains the driving force for Bose-Einstein condensation.

\begin{figure}[t]
\centerline{ \resizebox{6cm}{!}{ \includegraphics*{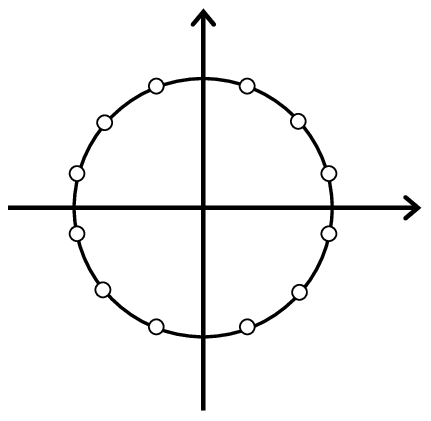} } }
\caption{\label{Fig:unit}
Cancelation of a permutation loop with an exponent that is not small.
Infinitesimal changes in position or momentum
give individual $\eta^{(l)}$ that are randomly and uniformly
distributed around the unit circle in the complex plane.
}
\end{figure}

Only permutations between bosons in the same momentum state
contribute to the occupation entropy,
but nevertheless permutations between bosons in different states
contribute to the symmetrized wave function.
Such permutations would have the superposed form of Eq.~(2)
if ${\bf p}''$ was a permutation of ${\bf p}'$,
and they are therefore absent from the classical world.
To explain how they are suppressed,
we focus on the permutation loops
into which all permutations may be factored.
The symmetrization function for the loop
given by a cyclic permutation of length $l$ is
\begin{eqnarray}
\eta^{(l)}_{{\bf p}^l}({\bf q}^l)
=
e^{-[{\bf p}-\hat{\mathrm P}^l_0{\bf p}]\cdot{\bf q}/\mathrm{i}\hbar}
=
e^{-{\bf p}_{j_1,j_2} \cdot {\bf q}_{j_1} /\mathrm{i}\hbar}
e^{-{\bf p}_{j_2,j_3} \cdot {\bf q}_{j_2} /\mathrm{i}\hbar}
\ldots
e^{-{\bf p}_{j_l,j_1} \cdot {\bf q}_{j_l} /\mathrm{i}\hbar} .
\end{eqnarray}
This is written in the form of a momentum loop;
swapping the positions with the momenta
is numerically equal and called the position loop form.
This loop symmetrization function
is a highly oscillatory function of the positions and momenta
that lies on the unit circle in the complex plane.
It therefore mostly cancels
upon averaging over small regions of phase space (Fig.~\ref{Fig:unit}).
In order to survive, the exponent must be small or zero,
in which case the symmetrization loop function is real and equal to unity.
A permutation is negligible unless all of its loop factors
have small or zero exponent.

For momentum loops, practically
this means that all bosons in the loop must be in the same momentum state,
${\bf p}_{j_l} = {\bf p}_{j_2} = \ldots  = {\bf p}_{j_l}$.
For position loops, it means that consecutive bosons in the loop
are separated by less than about the thermal wavelength,
which configurations are significant on the high temperature side
of the transition.
On the low temperature side momentum loops dominate,
and the only permitted permutations are the $\chi_{\bf p}^+$ permutations
between bosons in the same momentum state.
Such permutations are classically acceptable
because they do not superpose distinct states
(eg.\ Eq.~(2) with ${\bf p}' = {\bf p}''$).
Thus on the quantum side of the transition the permitted permutations
leave the momentum configuration of the subsystem unchanged,
$\hat{\mathrm P}{\bf p} = {\bf p}$,
which means that a validly symmetrized wave function
is the same as the unsymmetrized wave function,
$\zeta^+_{\bf p}({\bf q}) = \zeta_{\bf p}({\bf q})$.

Since the allowed permutations 
are those between bosons in the same momentum state,
the difference between the quantum and classical regimes
turns on whether the states are singly- or multiply-occupied.
If the momentum states are empty or at most singly-occupied,
then the subsystem behaves classically
because the occupation entropy is zero.
However, if some momentum states are multiply-occupied,
then the system has a quantum contribution to its properties and behavior,
depending upon the extent of condensation.
In general,
the greater the temperature, the greater the probability
that an individual boson will reach a high kinetic energy,
and therefore the greater the number of accessible momentum states.
At high temperatures and low densities,
there are many more accessible momentum states than there are particles,
and so the states are at most singly-occupied,
the occupation entropy is zero,
and the system behaves classically.
As the temperature is lowered,
the number of accessible momentum states decreases
until it becomes comparable to the number of bosons.
At this point there is substantial condensation
(ie.\ multiple occupancy of multiple momentum states)
and there is identifiable non-classical behavior.

\subsection{Non-Locality}

Non-locality is a feature of the quantum world
that makes it appear quite weird from the classical perspective.
Examples of non-locality include:
the quantization of states,
which is due to the way the boundaries, no matter how remote,
determine the allowed wave functions throughout the system;
the multiple occupancy of single-particle states
and the consequent occupation entropy,
which occurs irrespective of the location of the particles in the state;
and the entanglement and collapse of the wave function
instantaneously over macroscopic distances.

In general the spacing between quantized states
is inversely proportional to the size of system.
Hence for a macroscopic system the discrete states
go over to the classical continuum,
and the effects of quantization are generally immeasurably small.

As mentioned above the distinction between quantum and classical behavior
lies in part in whether the quantum states are multiply-occupied or not.
In the classical regime the states are empty or at most singly-occupied,
and so this aspect of quantum non-locality has no effect.
In the vicinity of the quantum-classical transition,
and in the quantum regime,
the effects of multiple occupancy are felt non-locally
and are manifest in such phenomena as superfluidity, as we shall see.

As also mentioned,
the entanglement of the subsystem with its environment
by energy conservation
causes the subsystem to collapse into decoherence.
The classical world as we perceive it
is a direct consequence of non-local entanglement,
even though classical phenomena are strictly local.

%
\section{Motion in Quantum and Classical Systems}
\setcounter{subsubsection}{0} 
%

\subsection{Equations of Motion}

In the uncondensed regime
Hamilton's classical equations of motion follow from Schr\"odinger's equation
for an open quantum system
by noting the absence of superposition states.
This is an empirical fact in the classical world:
a single position-momentum configuration in phase space
evolves to a new configuration, also unique.
Since the momentum eigenfunctions
reflect the positions and momenta of the particles,
this means that in an open quantum system
Schr\"odinger's time propagator over a short interval $\tau$ must yield
\begin{equation}
\left[ \hat{\mathrm I}
+ \frac{\tau}{\mathrm{i}\hbar}  \hat{\cal H}({\bf q})\right]
\zeta_{\bf p}({\bf q})
=
\zeta_{{\bf p}'}({\bf q}').
\end{equation}
Demanding time reversibility and continuity,
and with the relation between the Hamiltonian operator
and the classical Hamiltonian being
$\hat{\cal H}({\bf q}) \zeta_{\bf p}({\bf q})
= {\cal H}({\bf q},{\bf p})  \zeta_{\bf p}({\bf q})$,
this non-linear equation has solution
\begin{equation}
{\bf q}'
 =
{\bf q} + \tau \nabla_p {\cal H}({\bf q},{\bf p}),
\mbox{ and }
{\bf p}'
=
{\bf p} - \tau \nabla_q {\cal H}({\bf q},{\bf p}).
\end{equation}
These are Hamilton's classical equations of motion.
The second says that
the rate of change of momentum equals the force.
The equations conserve energy and hence reservoir entropy.
Because we are in the uncondensed regime
with each  momentum state either empty or singly occupied,
there is no contribution from the occupation entropy, $\chi_{\bf p}^+ = 1$.

In the condensed regime,
similar analysis
for a specific subset $A$ of $n_A$ bosons
in a momentum state ${\bf a}$ containing $N_{\bf a}$ bosons
gives
\begin{equation}
{\bf p}'^{n_A}
=
{\bf p}^{n_A}
+ \tau \frac{(N_{\bf a}-n_A)!n_A!}{N_{\bf a}!}  {\bf F}_{A}^{n_A}.
\end{equation}
The shared non-local force,
${\bf F}_A = n_A^{-1} \sum_{j \in A} {\bf f}_j$,
arises from permutations.
The binomial coefficient, 
which is the number of superposed possible subsets, 
reduces the force
so that the changes in the total kinetic and potential energies cancel.
These superposed momenta are suppressed
by entanglement with the environment,
to which the kinetic energy that they contain is dissipated,
leaving the specific subset $A$
and its reduced change in total kinetic energy (Attard 2024).

These forms of the momentum evolution invoke the thermodynamic limit
in which the spacing
between the momentum states goes to zero,
and the momentum is treated as a continuum.
In the condensed regime it is better to retain the quantization of momentum
and deal with the occupation of states directly.
In this case the equations of motion involve stochastic transitions
between the quantized momentum states,
and for these we shall need the probability distribution.

\subsection{Equilibrium Probability}

The decoherence of an open quantum system
as the origin of the classical world
was discussed above in terms of quantum statistical mechanics.
In this case the microstates are a momentum-position configuration,
${\bf \Gamma} = \{{\bf q},{\bf p}\}$,
a point in classical phase space but with quantized momentum.
For an equilibrium system
the probability of such a configuration is
\begin{equation}
\wp({\bf \Gamma})
= \frac{1}{Z} e^{-\beta {\cal H}({\bf \Gamma} )} \chi^+_{\bf p} ,
\end{equation}
where $Z$ is the normalizing partition function.
This comes from Eq.~(3) for fixed $N$
with the short-ranged commutation function neglected
(because condensation is non-local and driven by long-range effects)
and momentum loops dominating the symmetrization function,
$\eta^+({\bf \Gamma}) = \chi_{\bf p}^+$.
This goes over to the classical Maxwell-Boltzmann
probability in the limit of vanishing condensation,
$\chi^+_{\bf p} \to 1$.

In the high-temperature neighborhood
of the Bose-Einstein condensation transition,
coupling between the positional structure and
the sums or integrals over momenta of position loops
precludes the present factorization
wherein the symmetrization function is replaced
by the symmetrization factor
that depends only on the momentum configuration.

\subsection{Transition Rate}

For an open quantum system
the environment provides an outside influence
that can be characterized on average but not in detail,
which means that the time evolution of the subsystem
quantized momentum configurations
must be stochastic and given by a transition probability.
(The position evolution over a short time interval is deterministic,
${\bf q}' = {\bf q} + \tau {\bf p}/m$.)
At each instant in time an open quantum system
exists in a unique position-momentum configuration,
and the adiabatic conditional transition probability
obeys microscopic reversibility,
$\wp({\bf \Gamma}'|{\bf \Gamma},\tau) \wp({\bf \Gamma})
=
\wp({\bf \Gamma}^\dag|{\bf \Gamma}'^\dag,\tau) \wp({\bf \Gamma}'^\dag)$,
where the conjugate configuration with all the momenta reversed is
${\bf \Gamma}^\dag = \{{\bf q},-{\bf p}\}$,
and
$\wp({\bf \Gamma}^\dag) = \wp({\bf \Gamma})$.
This is satisfied by
the conditional transition probability
for $n_A$ bosons comprising a loop $A$
in a random permutation of the $N_{\bf a}$ bosons
in the momentum state ${\bf a}$
in time $\tau$
to the neighboring momentum state
${\bf a}'_\alpha = {\bf a} + \mbox{sign}(\tau F_{A,\alpha})\Delta_p$
in the direction of their shared non-local force
(Attard 2024),
\begin{equation}
\wp({\bf a}'_\alpha|{\bf a};n_A,\tau)
=
\frac{|\tau F_{A,\alpha}|}{\Delta_p}
\frac{ n_A! (N_{\bf a}-n_A)!}{N_{\bf a}!}
\left\{ 1
- \frac{\beta n_A \Delta_p}{2m}
\mbox{sign}(\tau {F}_{A,\alpha}) a_\alpha
+ \frac{\beta \tau a_\alpha}{2m} n_A {F}_{A,\alpha}
\right\} .
\end{equation}
The binomial coefficient is essential
for this to satisfy microscopic reversibility to first order
in $\tau$ and $\Delta_p$.
This 
probability gives the  transition rate
${\bf a} \stackrel{n_A,\tau}{\to} {\bf a}'_\alpha$
sequentially for  each neighboring destination ${\bf a}'_\alpha$,
each subset $A$,
and each state ${\bf a}$.

These motion 
on average conserves the total entropy, 
which is a general statistical requirement for an equilibrium system.
Experimental measurement of the fountain pressure
leads quantitatively to the thermodynamic principle
that superfluid flow minimizes energy at constant entropy
(Attard 2025 Ch.~4).
Thus there is consonance between the present stochastic equations of motion,
the general principles of equilibrium statistical mechanics,
and empirical superfluid thermodynamics.

In the classical regime, $N_{\bf a} = n_A =1 $,
the binomial coefficient is unity and
the prefactor ensures that Newton's second law of motion is obeyed:
the average rate of change of momentum is equal to the applied force.
In the quantum regime Newton's second law does not hold:
condensation means that
the average rate of change of momentum
for individual bosons, $n_A=1$,
is reduced from the classical result by a factor $ N_{\bf a}^{-1} $.
Since the occupation number of the low-lying momentum states
in the condensed regime is typically $ N_{\bf a} \sim {\cal O}(10^2)$,
this is a substantial reduction.
Transitions of larger subsets of bosons
are exponentially smaller;
the greatest reduction occurs when $n_A = N_{\bf a}/2$,
in which case it is $ {\cal O}( 2^{-N_{\bf a}})$.
Transitions of the momentum state as a whole, $n_A=N_{\bf a}$,
have no reduction in rate due to the  binomial coefficient,
but they comprise a tiny minority of possible transitions.

Further reduction in the rate of change of momentum
is due to the sharing of the force non-locally amongst
the bosons in the subset,
$ {\bf F}_A = n_A^{-1} \sum_{j\in A} {\bf f}_j$.
This shared non-local  force is a consequence of the permutations
of the bosons in the momentum state.
The magnitude of the change due to this for the momentum state
scales with $n_A^{1/2}$,
whereas the dissipative effects on the classical shear viscosity
of individual momentum changes scale with $n_A$.
This effect is largest for the largest subsets.
This non-local sharing effect
reduces the rate of change of momentum in shear flow
from its classical value.
The combinatoral effect of the occupation entropy
and the effect of the non-local sharing of the forces
explain at the molecular level
the reduction in viscosity in the condensed superfluid.

%
\section{The $\lambda$-Transition and Superfluidity}
\setcounter{subsubsection}{0} 
%

\begin{figure}[t!]
\centerline{ \resizebox{8cm}{!}{ \includegraphics*{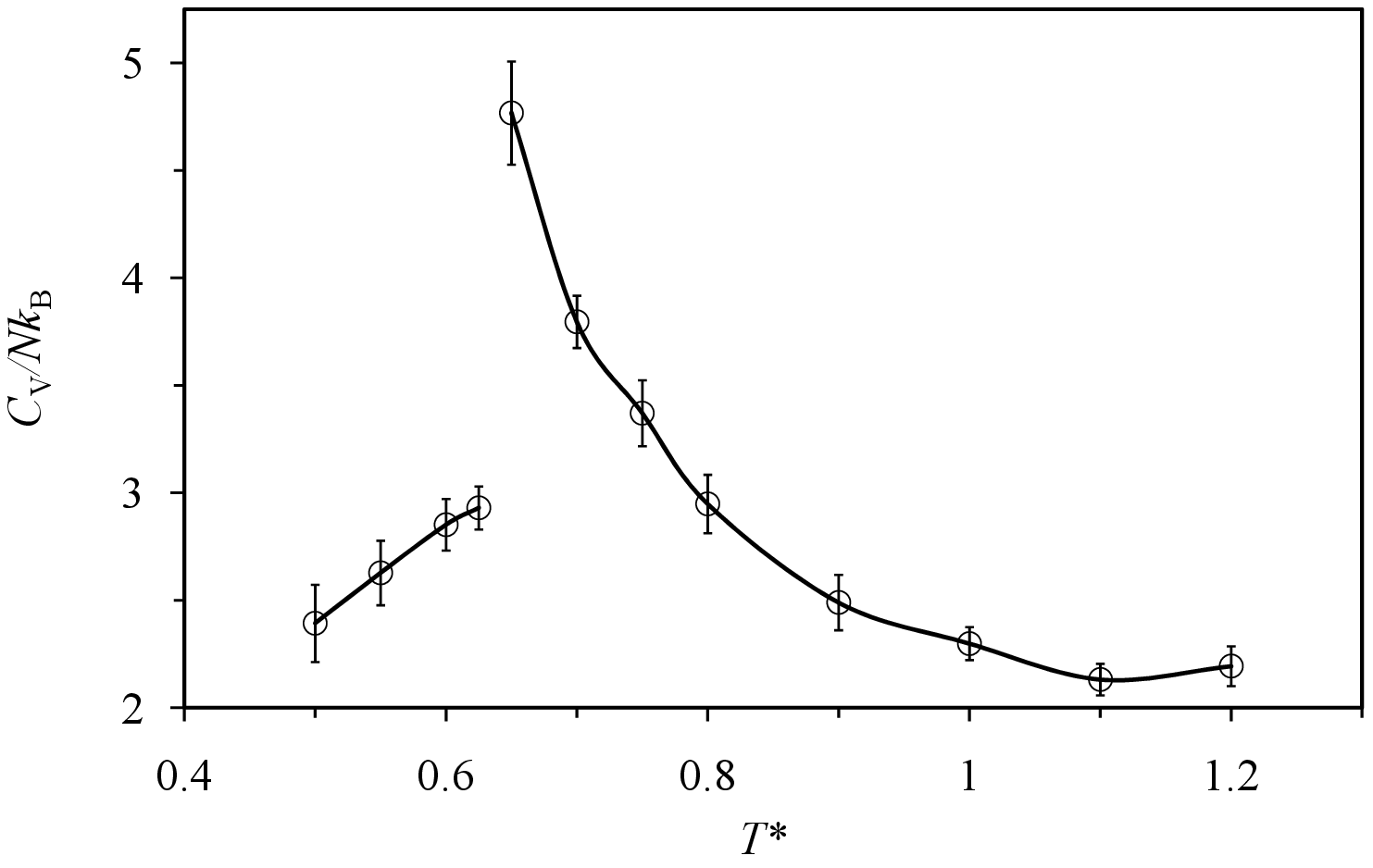} } }
\caption{
The $\lambda$-transition in the specific heat
in saturated Lennard-Jones $^4$He
by quantum particle Monte Carlo simulation
using up to pentamer position permutation loops above the transition
($T[K] = 10.22T^*$).
Below the transition
pure momentum permutation loops are used.
From Attard (2025 Fig.~3.5).
}\label{Fig:CvLJ}
\end{figure}

Figure~\ref{Fig:CvLJ}
shows the heat capacity obtained
with quantum particle Monte Carlo simulations
of $^4$He atoms interacting with the Lennard-Jones pair potential.
Equation~(3) with fixed number was used for the probability,
with the commutation function neglected.
On the high temperature side of the transition
position permutation loops were included
up to pentamers
in an exponential resummation of the symmetrization function
(Attard 2018, 2021, 2025).
The growth in size and number of these causes the divergence
in the heat capacity. 
Position loops and momentum loops are mutually exclusive;
each dominate their respective side of the transition.
On the low temperature side only momentum permutation loops were used.
In this regime
the decline in the heat capacity with decreasing temperature
is due to the increasing occupancy of low-lying momentum states:
as the kinetic energy decreases,
so also the magnitude of its fluctuations.
The liquid structure changes little with temperature
and hence the contribution of the potential energy
to the heat capacity is relatively small.

It is worth critically analyzing the known experimental evidence
for the $\lambda$-transition
in order to understand the precise nature of Bose-Einstein condensation.
There are three empirical facts:
at the $\lambda$-transition
(1) there is no latent heat,
(2) the slope of the heat capacity is discontinuous,
and (3) superfluidity is discontinuous.
If we accept that the  $\lambda$-transition signifies
the onset of Bose-Einstein condensation,
then from the second and third points we conclude
that the change in condensation is discontinuous and macroscopic.
From the third point we conclude
that the condensation is zero above the transition.
From the first point we conclude that condensation
is not solely into the ground state.
This follows because a macroscopic number of bosons
discontinuously occupying the ground state from excited states
would give an energy discontinuity and a latent heat.
Rather, condensation must be into multiple low-lying states,
with highly and lowly occupied states intercalated.
The total kinetic energy in the states
in any finite energy or momentum interval must be more or less unchanged
by the transition,
which means that the bosons in highly occupied states
must come from nearby states that are now low-occupied.
This is consistent with the analysis of Bose-Einstein condensation
for ideal bosons,
namely that in the condensed regime
the relative fluctuations in occupancy are order unity.

Thermodynamics and statistical mechanics show
that condensation into the ground momentum state is intensive,
with the average occupancy being ${\cal O}(10^2)$
and therefore on its own relatively negligible.
In the thermodynamic limit this number is about the same
as the average occupancy of any low-lying momentum state.
This is one reason for paying attention to the occupancy
of the ground momentum state even though condensation
is not solely into the ground momentum state.
After the condensation transition,
the instantaneous number of highly occupied low-lying momentum states
is macroscopic.

The $\lambda$-transition signifies the quantum-classical transition.
Since there is no latent heat at the transition,
the first temperature derivative of the free energy is continuous.
Hence it may be said that the  quantum-classical transition is continuous.

\begin{figure}[t]
\centerline{ \resizebox{8cm}{!}{ \includegraphics*{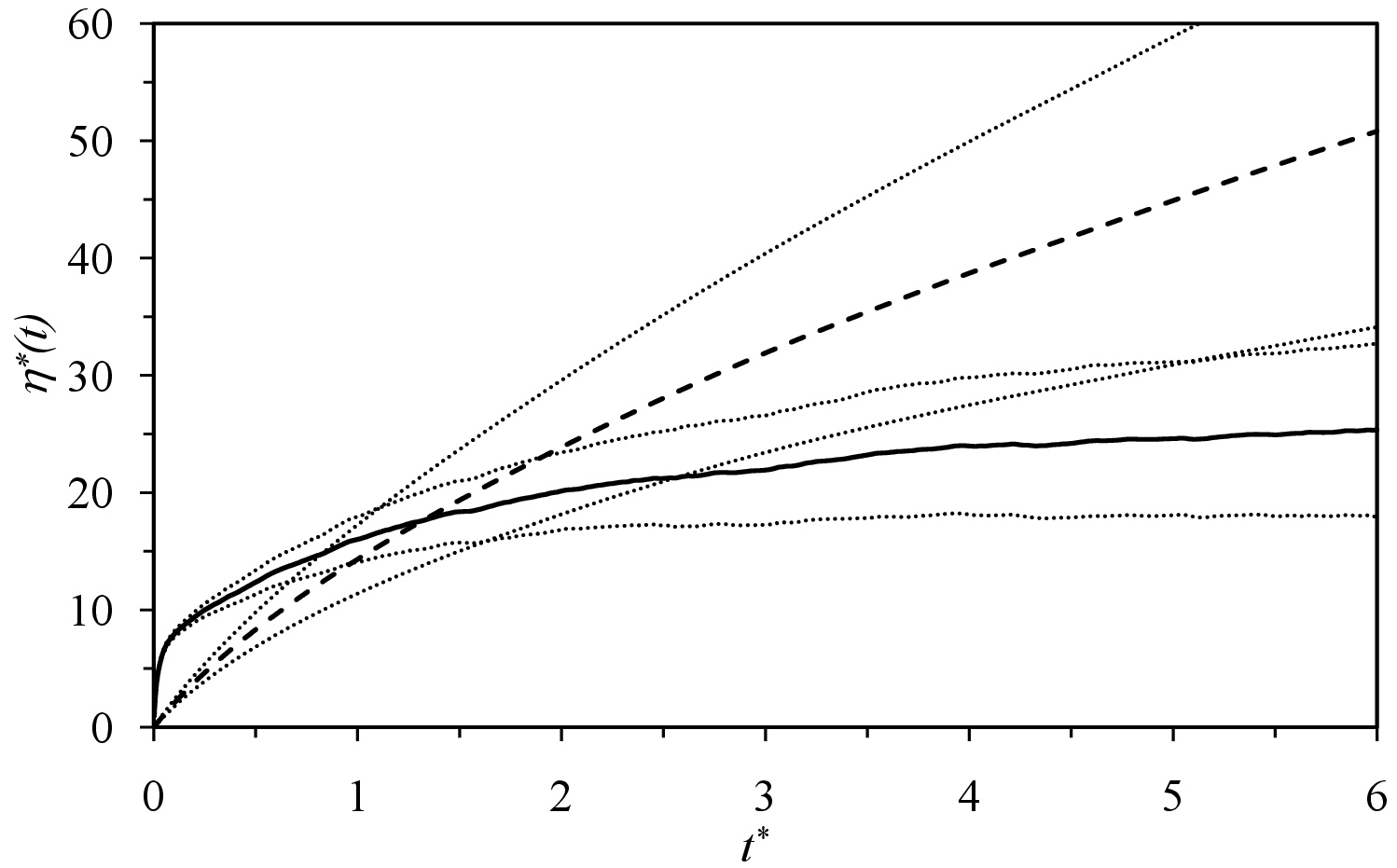} } }
\caption{\label{Fig:eta-pbc}
Viscosity time function
from quantum molecular dynamics simulations of liquid Lennard-Jones $^4$He
at $T^*=0.60$, $\rho^*=0.8872$ ($N=1,000$).
The solid curve is the quantum liquid,
the dashed curve is the classical liquid,
and the dotted curves give the 95\% confidence interval.
From Attard (2024).
}
\end{figure}

Figure~\ref{Fig:eta-pbc} compares the viscosity
for quantum and classical Lennard-Jones $^4$He.
The two simulations were identical
apart from the occupation entropy contribution 
being set to unity in the classical simulations
(ie.\ $n_A=N_{\bf a}=1$ in Eq.~(10)).
The maximum value of the viscosity time function
is `the' shear viscosity,
and for the quantum liquid this is $\eta^\mathrm{qu}(6)= 25.4(74)$.
This is about one quarter of the extrapolated maximum
for the classical liquid, $\eta^\mathrm{cl}(20.2)= 91(53)$.

Conceptually we can make a binary division into uncondensed bosons,
which are those in singly occupied momentum states,
and condensed bosons,
which are those in momentum states with two or more.
With this definition,
the simulations give about three quarters of the bosons as condensed
in the quantum case, $f_0^\mathrm{qu} = 0.7473(4)$,
compared to $f_0^\mathrm{cl} = 0.50713(7)$ in the classical case.
In contrast,
the conventional two-fluid model of superfluidity
makes the binary division into ground state bosons
and excited state bosons,
with the former responsible for superfluidity.
It models the total viscosity
as the fraction of excited state bosons times their ordinary viscosity,
ascribing zero viscosity to bosons in the ground state.
Instead, identifying superfluid bosons with the present condensed bosons,
and the excited state viscosity with the classical viscosity,
the  linear binary model would give
$\eta^\mathrm{qu} = (1-f_0^\mathrm{qu})\eta^\mathrm{cl}$.
This is in surprisingly good agreement with the  simulated value above
(and also others at higher temperatures).
Given the simplicity of the model,
the arbitrary binary definition of condensation,
and the statistical error and uncertainty in the classical viscosity,
not too much should be read into this.

Rather than confirming the original two-fluid model,
the present definition of condensation
is in fact antithetical to ground state occupation.
In this context, `measured' experimental values in the literature
for the fraction of ground state bosons
that are based on the original two-fluid model should be treated skeptically.
This view is reinforced by the nature of the $\lambda$-transition
as discussed following Fig.~\ref{Fig:CvLJ}:
condensation into the ground state cannot be macroscopic.

With $N=1,000$,
the average occupancy of the ground momentum state
in the quantum case is $\overline N^\mathrm{qu}_{000} = 139.6(90)$.
(In the classical case it is $\overline N^\mathrm{cl}_{000} = 2.438(5)$.)
The occupancy is an intensive variable
that changes slowly with system size,
and so in the thermodynamic limit
$\overline N_{\bf n} = {\cal O}(10^2)$
can be expected to be representative
of the average occupancy of each of the low-lying momentum states.

The present simulations were performed
for a homogeneous system with periodic boundary conditions.
Different versions of the conditional transition probability, Eq.~(10)
(eg.\ ${\bf F}_A \Rightarrow {\bf F}_{\bf a}$,
or $n_A! \Rightarrow 1 \mbox{ or } \delta_{n_A,1}$),
were also explored,
and also an inhomogeneous system with uniform Lennard-Jones walls
(Attard 2024).
The quantitative value of the quantum viscosity
was rather insensitive to the details
of the conditional transition probability, system size, and geometry,
and in all cases in the condensed regime
it was a fraction of the classical viscosity.
With increasing temperature
the quantum viscosity,
the fraction of condensed bosons,
and the ground state occupancy
all approach their classical values.
Because position permutation loops are not included
in the quantum molecular dynamics simulations,
there is no discontinuous transition to condensation or to superfluidity,
which is in contrast to the quantum  particle Monte Carlo  simulations of
Fig.~\ref{Fig:CvLJ}.

It was mentioned above
that quantum non-locality plays a role in superfluidity.
The quantization of the momentum states
and the subsequent well-defined occupancies
are due to the macroscopically remote boundaries of the system.
Bosons in the same momentum state are tied together
by the occupation entropy
irrespective of where they are located in the system.
The equations of motion, Eqs~(8) and (10),
show how the rate of change of momentum of such condensed bosons
is reduced from the classical rate.
Whereas the shear flow of a classical viscous fluid
is necessarily organized in space,
superfluid flow is akin to the plug flow of an inviscid fluid
precisely because of the non-local nature
of the occupation of the momentum states.

\section{Conclusion}

Understanding the quantum-classical transition relies on several ideas:
(1) that particles really have position and momentum and follow trajectories,
(2) that superposition states cancel in an open quantum system,
and (3) that full symmetrization of the wave function
is unnecessary because most permutations cancel,
the exceptions being those amongst bosons
either in the same momentum state
or else close together in position space.
In consequence of this third point,
Newton's second law of motion is valid if the
momentum states are empty or singly occupied,
but it overestimates the rate of change of momentum
due to an applied force in the condensed regime.
Additionally, (4) quantum non-locality is manifest
in Bose-Einstein condensation and superfluidity.

The $\lambda$-transition in liquid helium
is due to Bose-Einstein condensation,
and as such it may be considered a manifestation
of the quantum-classical transition.
It is continuous,
which justifies the formulation of the evolution of a quantum system
in terms akin to the classical equations of motion for particles.
The absence of viscosity in superfluid flow
ultimately arises from the preservation of occupation entropy,
which has the consequence
that the rate of change of momentum due to an applied force
decreases exponentially
with increasing occupation number in the condensed regime.

\section*{References}


\begin{list}{}{\itemindent=-0.5cm \parsep=.5mm \itemsep=.5mm}

\item 
Attard P 2018
Quantum statistical mechanics in classical phase space. Expressions for
the multi-particle density, the average energy, and the virial pressure
arXiv:1811.00730

\item 
Attard P 2021
\emph{Quantum Statistical Mechanics in Classical Phase Space}
(Bristol: IOP Publishing)

\item 
Attard P 2024
The molecular nature of superfluidity: Viscosity of helium from quantum
  stochastic molecular dynamics simulations over real trajectories
arXiv:2409.19036v5

\item 
Attard P 2025
\emph{Understanding Bose-Einstein Condensation,
Superfluidity, and High Temperature Superconductivity}
(London: CRC Press)

\item 
Bohm D 1952
A suggested interpretation of the quantum theory
in terms of `hidden' variables. I and II.
\emph{Phys.\ Rev.}\ {\bf 85} 166 


\item 
de Broglie L 1928
La nouvelle dynamique des quanta,
in Solvay   p.~105

\item 
Goldstein S 2024
Bohmian mechanics
\emph{The Stanford Encyclopedia of Philosophy}
(Summer 2024 Edition), E N Zalta and U  Nodelman (eds.)
{\tt URL =
<https://plato.stanford.edu/archives/\\
sum2024/entries/qm-bohm/>}

\item 
Joos E and Zeh H D 1985
The emergence of classical properties through
interaction with the environment
\emph{Z.\ Phys.}\  B {\bf 59} 223


\item 
Schlosshauer M 2005
Decoherence, the measurement problem,
and interpretations of quantum mechanics
arXiv:quant-ph/0312059v4

\item 
Zurek W H 1991
Decoherence and the transition from quantum to classical
\emph{Phys.\ Today} {\bf 44} 36

\end{list}

\end{document}